\documentclass[doublecol]{epl2}
\usepackage{graphicx}
\usepackage[utf8,latin1]{inputenc}
\usepackage{dcolumn}
\usepackage{amsmath}
\usepackage{amssymb}
\usepackage[square,comma,numbers]{natbib}
\newcommand{\qm}[1]{``#1''}

\title{Analytical coordinate time at first post-Newtonian order}
\shorttitle{Analytical coordinate time at first post-Newtonian order}

\author{Vittorio De Falco$^{1,2}$\thanks{vittorio.defalco-ssm@unina.it} \and Emmanuele Battista$^{3}$\thanks{emmanuele.battista@univie.ac.at,\\ emmanuelebattista@gmail.com}\and John Antoniadis$^{4,5}$\thanks{john@ia.forth.gr}}
\shortauthor{De Falco, Battista, Antoniadis (2022)}

\institute{$^1$ Scuola Superiore Meridionale, Largo San Marcellino 10, 80138 Napoli, Italy,\\
$^2$ Istituto Nazionale di Fisica Nucleare, Sezione di Napoli, Complesso Universitario di Monte S. Angelo, Via Cintia Edificio 6, 80126 Napoli, Italy\\
$^3$ Department of Physics, University of Vienna, Boltzmanngasse 5, A-1090 Vienna, Austria\\
$^4$ Institute of Astrophysics, FORTH, Dept. of Physics, University Campus, GR-71003 Heraklion, Greece \\
$^5$ Max-Planck-Instutut f\"ur Radioastronomie, Auf dem H\"ugel 69, 53121, Bonn, DE}

 
\abstract{In this letter, we exploit the Damour-Deruelle solution to derive the  analytical expression of the coordinate time in terms of the polar angle. This formula has advantageous applications in both pulsar timing and gravitational-wave theory. }

\begin{document}

\maketitle

\section{Introduction} 
During the past few decades, General Relativity (GR) has been subjected to a number of experiments beyond the weak-field regime of the Solar System. A crucial astrophysical testbed of GR is represented by compact binary systems. However, due to the non-linear geometric structure of GR, the dynamics of a binary system is contained in the gravitational field equations and it is governed by retarded-partial-integro differential equations  \cite{Maggiore:GWs_Vol1,Blanchet2014,Poisson-Will2014}. These issues complicate subsequent calculations, making it difficult to test GR against observations.

The aforementioned criticality has been solved via approximation strategies and by adopting special coordinate systems (e.g., harmonic coordinates), at the price of losing the general covariance of the GR theory. Furthermore, the gravitational source is assumed to be post-Newtonian (PN), i.e., slowly moving, weakly self-gravitating, and weakly stressed \cite{Maggiore:GWs_Vol1,Blanchet2014}. This allows us to apply in the near zone the PN approximation method \cite{Lorentz1937}, which produces instantaneous potentials with no retardation effects \cite{Maggiore:GWs_Vol1,Blanchet2014}. In addition, assuming that the bodies are well separated, we can deal with test particles and ordinary differential equations \cite{Blanchet2014,Poisson-Will2014}. The bodies' motion occurs in the Newtonian absolute Euclidean space, in which we add the PN corrections. Moreover, these dynamical equations preserve their relativistic nature, since they remain invariant under a global PN-expanded Lorentz transformation \cite{Blanchet2014}.

The PN method has been extensively employed to investigate several aspects of the relativistic two-body dynamics, including the back-reaction of gravitational radiation in binary pulsars \cite{Taylor:1989sw,Damour:1990wz,Kramer:2006nb,Kramer:2021jcw} and the direct detection of gravitational waves (GWs) from coalescing compact binaries \cite{Blanchet2014}. The PN formalism has also played a central role in precision tests of gravity theories \cite{Wex:2014abc} and neutron star mass measurements in binary pulsars \cite{Ozel:2016oaf}. These assessments extensively use the quasi-Keplerian analytical solution of Damour and Deruelle, describing the quasi-elliptic 1PN-accurate GR motion of a two-body system \cite{Damour1985}. 

In this letter, we start from the latter reference to derive, for the very first time, an analytical expression of the coordinate time, $t$, in terms of the polar angle, $\varphi$, at the 1PN level, namely the function $t=t(\varphi)$. The motivation for this work takes its origin from the need to speed up our numerical simulations for computing the gravitational waveforms and the fluxes from inspiralling binaries framed in GR (and, as recently proposed in the literature, also in Einstein-Cartan theory \cite{Paper2}). As the two-body dynamics can be readily analysed via the relative distance $R=R(\varphi)$  \cite{Damour1985}, then all dynamical quantities can depend on $\varphi$ and hence it is natural to invest efforts for determining $t(\varphi)$. 

This analytical formula can be used to replace numerical derivation schemes currently exploited in pulsar timing software such as \texttt{TEMPO} \cite{Taylor:1989sw}, \texttt{TEMPO2} \cite{Edwards:2006zg}, and \texttt{PINT} \cite{Luo:2020ksx}. Similarly, it may also be exploited in coherent pulsar search algorithms \cite{lentali2018,freire2018}, in which a very large number of trial timing solutions must be generated and compared with the data, in a manner similar to template matching in ground-based GW astronomy.

\section{Preliminaries} We consider a binary system composed of two PN weakly self-gravitating, slowly moving, and widely separated bodies with masses $m_1\ge m_2$, total mass $M:=m_1+m_2$, reduced mass $\mu:=\tfrac{m_1m_2}{M}$, symmetric mass ratio $\nu:=\tfrac{\mu}{M}$, position vectors $\boldsymbol{r}_1$, $\boldsymbol{r}_2$, and velocities $\boldsymbol{v}_1:=\frac{{\rm d}\boldsymbol{r}_1}{{\rm d}t}$, $\boldsymbol{v}_2:=\frac{{\rm d}\boldsymbol{r}_2}{{\rm d}t}$. 

\begin{figure}[h!]
    \centering
    \includegraphics[trim=4.5cm 3cm 3cm 0cm,scale=0.29]{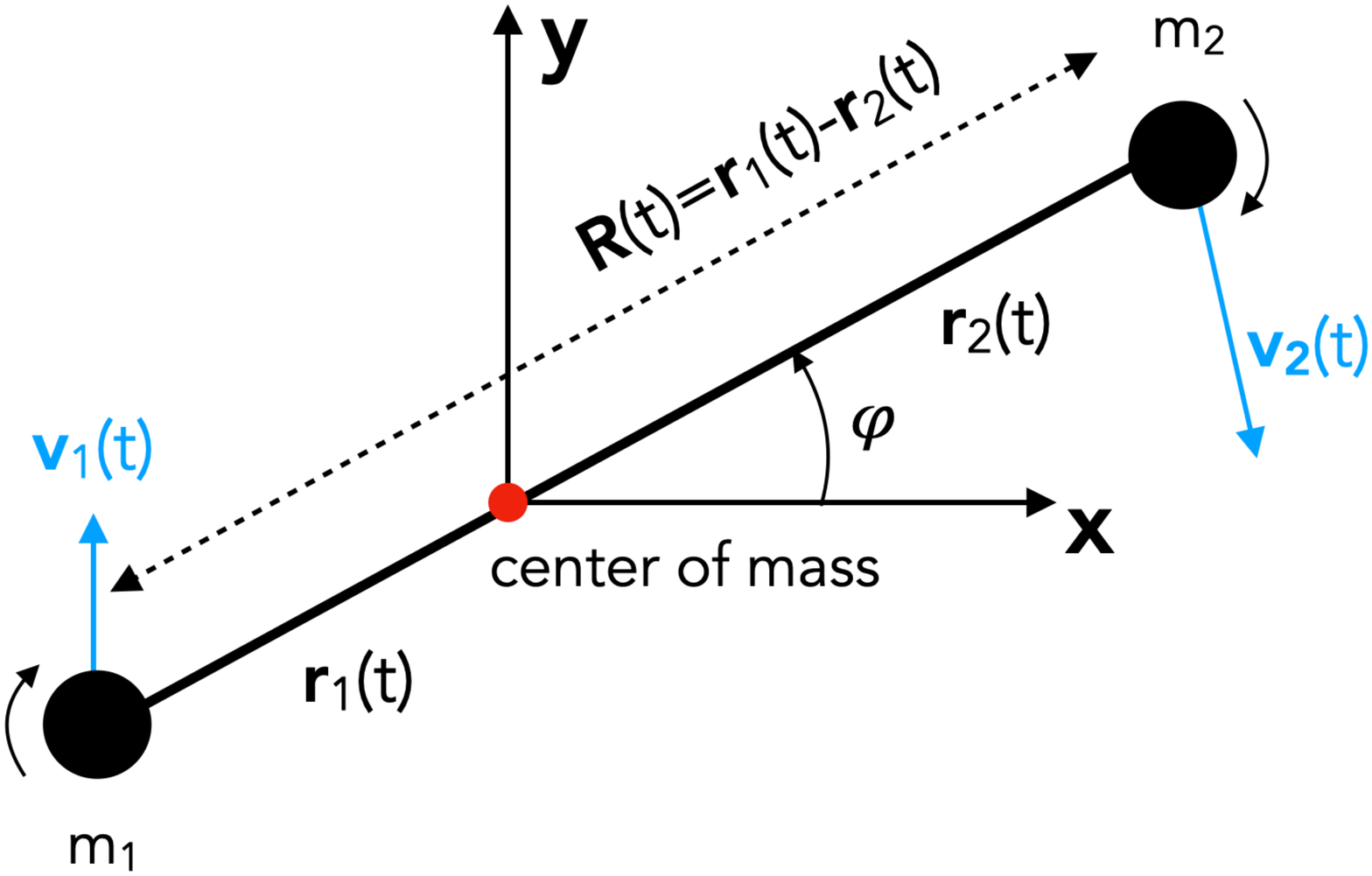}
    \caption{The barycentric coordinate system.}
    \label{fig:FigA0}
\end{figure}

Let us define the relative position, $\boldsymbol{R}:= \boldsymbol{r}_1(t)-\boldsymbol{r}_2(t)$, and the relative velocity, $\boldsymbol{V}:=\boldsymbol{v}_1(t)-\boldsymbol{v}_2(t)$, vectors. Then, likewise the Newtonian case, the description of the motion can be simplified by choosing  a fixed orthogonal reference frame $(\boldsymbol{x},\boldsymbol{y})$ centered in the barycenter of the binary system, which, without loss of generality, is supposed to be static. We also introduce $\varphi$ as the angle between $\boldsymbol{R}$ and the $\boldsymbol{x}$-axis, which is measured counterclockwise (see Fig. \ref{fig:FigA0}). In this frame, one can write
\begin{subequations}
\begin{align}
\boldsymbol{r}_1(t)&=\frac{\mu}{m_1}\boldsymbol{R}(t)+\frac{\mu(m_1-m_2)}{2M^2c^2}\left(V^2-\frac{GM}{R}\right) \boldsymbol{R}(t)\notag\\
&+{\rm O}\left(c^{-4}\right),\\
\boldsymbol{r}_2(t)&=-\frac{\mu}{m_2}\boldsymbol{R}(t)+\frac{\mu(m_1-m_2)}{2M^2c^2}\left(V^2-\frac{GM}{R}\right) \boldsymbol{R}(t)\notag\\
&+{\rm O}\left(c^{-4}\right),
\end{align}
\end{subequations}
where $R := |\boldsymbol{R}|$ and $V:=|\boldsymbol{V}|$.

In the GR framework, we consider the 1PN-accurate \emph{Damour-Deruelle treatment}, where the total conserved energy $E=E_0+\frac{1}{c^2}E_1 +{\rm O}\left(c^{-4}\right)$ and the angular momentum $\boldsymbol{J}=\boldsymbol{J}_0+\frac{1}{c^2}\boldsymbol{J}_1 +{\rm O}\left(c^{-4}\right)$ of the system are \cite{Damour1985} 
\begin{subequations} \label{PN_expansion_1}
\begin{align}
E_0 &:=\frac{1}{2}V^2-\frac{GM}{R},\\
E_1&:=\frac{GM}{2R}\left[(3+\nu)V^2+\nu\left(\frac{\boldsymbol{R}\cdot\boldsymbol{V}}{R}\right)^2+\frac{GM}{R}\right]\notag\\
&+\frac{3}{8}V^4 (1-3\nu),\\
J_0&:=\vert \boldsymbol{J}_0\vert=\vert\boldsymbol{R}\times\boldsymbol{V}\vert,\\
J_1&:=\vert \boldsymbol{J}_1\vert=\vert\boldsymbol{R}\times\boldsymbol{V}\vert\left[\frac{(1-3\nu)}{2}V^2+(3+\nu)\frac{GM}{R}\right],\\
J&:=\vert \boldsymbol{J}\vert=J_0+\frac{1}{c^2}J_1,
\end{align}
\end{subequations}
where  $|\boldsymbol{R}\times\boldsymbol{V}|=R^2 \frac{{\rm d}\varphi}{{\rm d}t}$ and $(\boldsymbol{R}\cdot\boldsymbol{V})=\frac{{\rm d}R}{{\rm d}t}$. It is important to note that the first integrals \eqref{PN_expansion_1} can be easily calculated once the initial conditions are assigned. Therefore, at the initial time $t_{\rm in}$ we have
\begin{align}
R(t_{\rm in})&=R_{\rm in},\quad \frac{{\rm d}R}{{\rm d}t}(t_{\rm in})=\dot{R}_{\rm in}, \notag \\ 
\varphi(t_{\rm in})&=\varphi_{\rm in},\quad\frac{{\rm d}\varphi}{{\rm d}t}(t_{\rm in})=\beta \sqrt{\frac{GM}{R_{\rm in}^3}},    \end{align}
where $0<\beta\le 1$ with $\beta=1$ corresponding to circular orbits. Having defined the parameters
\begin{subequations} \label{PN_expansion_2}
\begin{align}
Q_0&:= \frac{2GM}{R_0},\\
h_0&:= \frac{J_0}{GM},
\\
e_0&:=\sqrt{1+2E_0 h_0^2}, \qquad (0\le e_0 < 1),
\label{eq:e0-parameter-orbit}
\\
W_0&:=\frac{J_1}{J_0}-\frac{3}{h_0^2},\\
\tilde{K}&=1-\frac{1}{c^2}\frac{3}{h_0^2},
\end{align}
\end{subequations} 
the 1PN expansion of the radius $R=R_0+\frac{1}{c^2}R_1 +{\rm O}\left(c^{-4}\right)$ reads as
\begin{subequations} \label{PN_expansion_3}
\begin{align}
R_0&:=\frac{h_0^2GM}{1+e_0 \cos \left[\left(\varphi -\varphi_{\rm in} \right)\tilde{K}\right]}\,
\label{eq:R0} ,\\
R_1&:=\frac{1}{2} \left\{G\mu-\frac{E_0^2R_0^2}{GMe_0} \cos \left[\left(\varphi -\varphi_{\rm in}\right)\tilde{K}\right]\biggr{[} \nu -15\right.\notag\\
&\left.\left.+\frac{4W_0}{E_0}+\frac{2 E_1}{E_0^2}\right]+4R_0 \left[\frac{E_0}{2} (\nu -4)+ W_0\right]\right\}. \label{eq:R1}
\end{align}
\end{subequations}
From the above equations, it is clear that the Damour and Deruelle solution links the relative radius $R$ with the polar angle $\varphi$.\\

\section{Analytical expression of time} We start from Eq. (2.16) in Ref. \cite{Damour1985}, which naturally expresses the time $t$ in terms of the polar angle $\varphi$ up to ${\rm O}(c^{-4})$ corrections,
\begin{equation} \label{eq:Time_original}
{\rm d}t=\frac{{\rm d}\varphi}{\frac{H}{R^2}+\frac{I}{R^3}},    
\end{equation}
where
\begin{subequations}
\begin{align}
H&=J\left[1+(3\nu-1)\frac{E}{c^2}\right],\\
I&=(2\nu-4)\frac{GMJ}{c^2}.
\end{align}
\end{subequations}
Substituting the PN expansions \eqref{PN_expansion_1}, \eqref{PN_expansion_2}, and \eqref{PN_expansion_3} into Eq.\,\eqref{eq:Time_original}, we  expand the ensuing expression in power series of $1/c\to 0$ and up to the first order. In this way, after a lengthy calculation, we obtain the PN-expanded differential equation for the coordinate time, which, up to ${\rm O}(c^{-4})$ terms, reads as
\begin{align} \label{eq:dt-equat-1PN}
{\rm d}t&=\frac{R_0^2}{J_0}\biggr{\{}1+\frac{1}{c^2 }\biggr{[} E_0(1-3 \nu ) +\frac{2R_1}{R_0}\notag\\
&\qquad\qquad- 2Q_0(\nu -2)-\frac{J_1}{J_0}\biggr{]}\biggr{\}}
{\rm d}\varphi.    
\end{align}
This amounts to solve the integral\footnote{Since the two-body equations of motion in GR constitutes an autonomous dynamical system (i.e., independent of the time $t$) \cite{Damour1985}, we can assume $t_{\rm in}=0$ without loss of generality.}
\begin{align} \label{eq:time_angle}
t=\int g(\tilde{K}\bar{\varphi}){\rm d}\bar{\varphi},   
\end{align}
where $\bar{\varphi}:=\varphi-\varphi_{\rm in}$ and
\begin{align}
g(\tilde{K}\bar{\varphi})&:=g_1(\tilde{K}\bar{\varphi})+g_2(\tilde{K}\bar{\varphi})+g_3(\tilde{K}\bar{\varphi}),
\end{align}
with 
\begin{subequations}
\label{eq:function-g1,g2,g3}
\begin{align}
g_1(\tilde{K}\bar{\varphi})&:=A_1 R_0,
\label{eq:function-g1}
\\
g_2(\tilde{K}\bar{\varphi})&:=A_2R_0^2,
\\
g_3(\tilde{K}\bar{\varphi})&:=A_3 \cos(\tilde{K}\bar{\varphi}) R_0^3.   
\label{eq:function-g3}
\end{align}
\end{subequations}
The coefficients $A_1,A_2,A_3$ occurring in Eq. \eqref{eq:function-g1,g2,g3} can be obtained by substituting the expression \eqref{eq:R1} into Eq. \eqref{eq:time_angle}. In this way, we end up with
\begin{subequations}
\begin{align}
A_1&:=\frac{4-\nu}{c^2h_0},\\
A_2&:=\frac{J_0 \left[c^2-E_0 (\nu +7)+4 W_0\right]-J_1}{c^2 J_0^2},\\
A_3&:=-\frac{E_0^2 (\nu -15)+4 E_0 W_0+2 E_1}{c^2  GM E_0 J_0 }.
\end{align}
\end{subequations}

\begin{figure}
    \centering
    \includegraphics[scale=0.29]{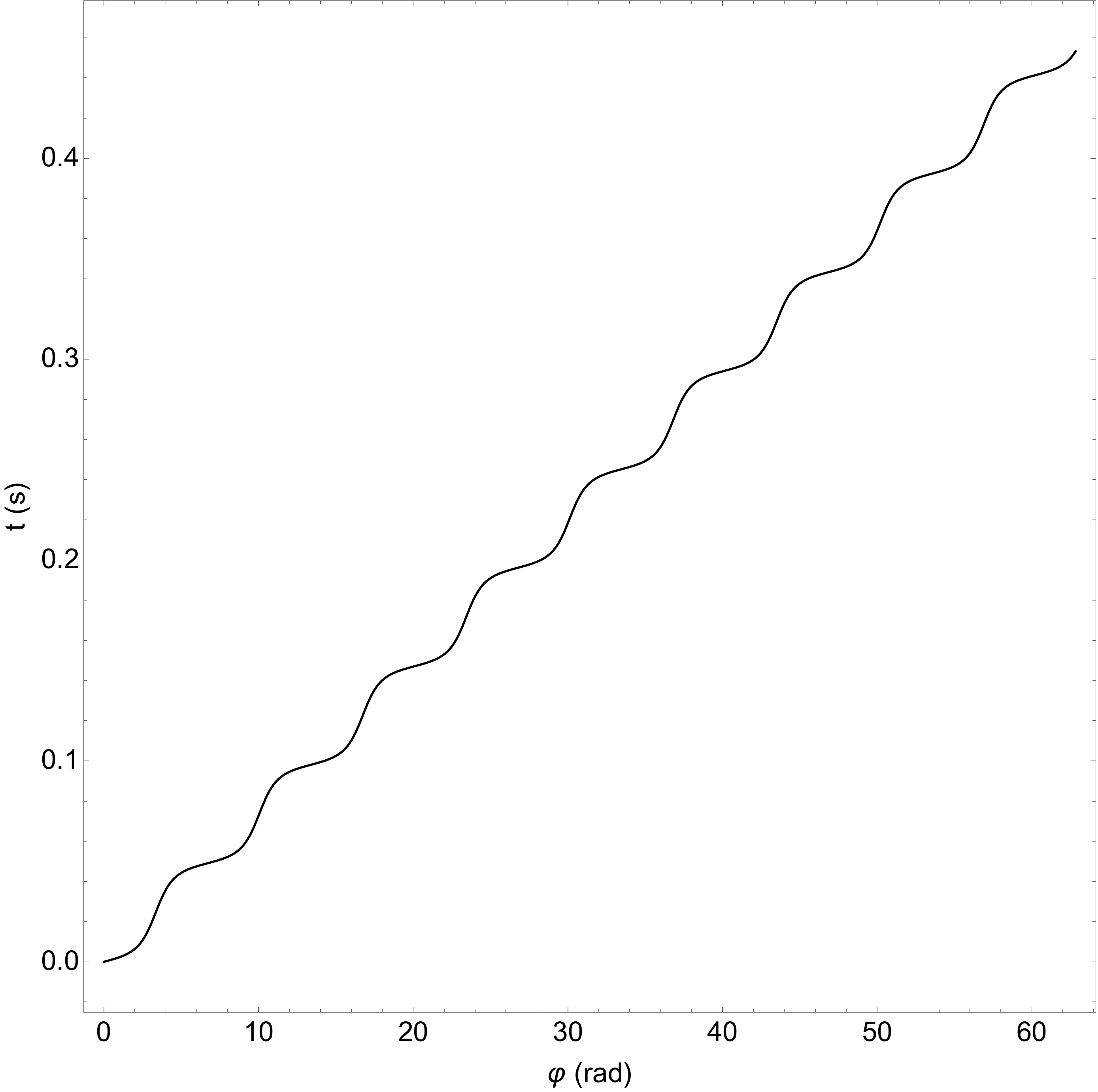}
    \caption{Numerical integration of $t(\varphi)$ for $\varphi\in[0,20\pi]$. The following values have been used: $m_1=1.60\ M_\odot$, $m_2=1.17\ M_\odot$, $R_{\rm in}=100 \frac{GM}{c^2}=410.83\ {\rm km}$, $\dot{R}_{\rm in}=0$, $\beta=0.7$,  $E_0=-6.80\times 10^{14}\ {\rm m^{2}\ s^{-2}}$, $E_1=1.07\times 10^{30}$, $J_0=8.62\times 10^{12}\ {\rm m^2\ s^{-1}}$, $J_1=2.57\times 10^{28}\ {\rm s}$.}
    \label{fig:FigA1}
\end{figure}

If we integrate Eq. \eqref{eq:time_angle} numerically, we get a \emph{monotonically increasing} function $t(\varphi)$ (see Fig. \ref{fig:FigA1}). 

\begin{figure}
    \centering
    \includegraphics[scale=0.29]{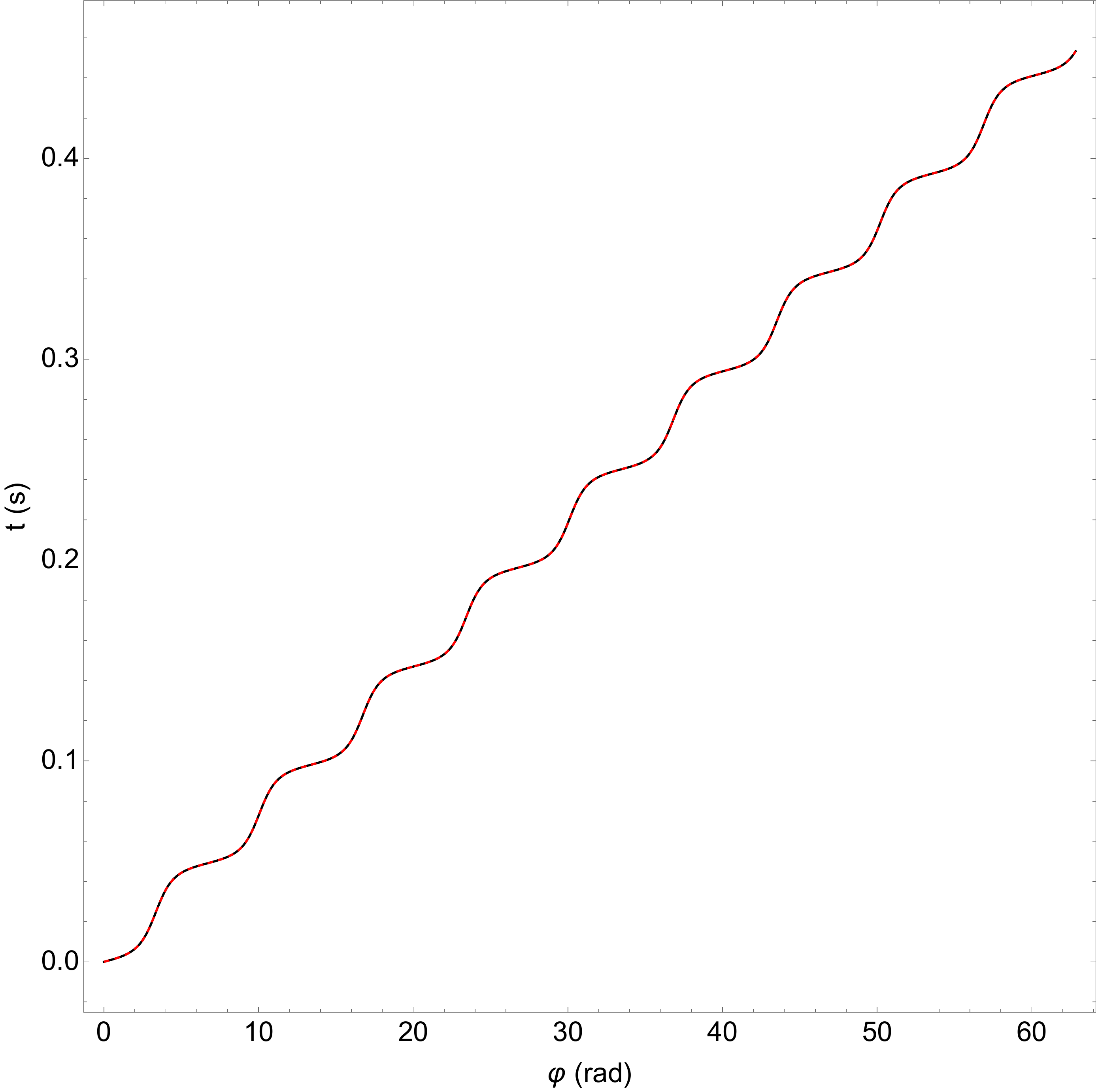}
    \caption{Function $t(\varphi)$ for $\varphi\in[0,20\pi]$, where the same numerical values as in Fig. \ref{fig:FigA1} are employed. The black continuous line represents the numerical solution, whereas the red dashed line the analytical expression \eqref{eq:anal_sol}.}
    \label{fig:FigA3}
\end{figure}

It is useful to write Eq. \eqref{eq:R0} equivalently as
\begin{align}
R_0=\frac{1}{B_1+B_2\cos(\tilde{K}\bar{\varphi})},
\end{align}
where
\begin{align} \label{eq:B1_B2}
B_1&:= \frac{1}{h_0^2 GM},\qquad B_2:= \frac{e_0}{h_0^2 GM},
\end{align}
with $B_1\ge B_2 \ge0$ (cf. Eq. \eqref{eq:e0-parameter-orbit}). Let us also define
\begin{align}
f_i(\varphi):= \int g_i(\tilde{K}\bar{\varphi}) {\rm d}\bar{\varphi},\quad i=1,2,3.    
\end{align}
The solution $t(\varphi)$ is obtained by adding the above three integrals (cf. Eqs. \eqref{eq:time_angle}--\eqref{eq:function-g1,g2,g3}), which we now solve separately.  For each integral, we use the same integration strategy: we first make 
the substitution $x=\tilde{K}\bar{\varphi}$ and then convert the trigonometric functions into polynomials through \emph{Weierstrass substitution} $\tau=\tan(x/2)$, leading to the following transformations:
\begin{align}
{\rm d}x&=\frac{2}{1+\tau^2}{\rm d}\tau,\qquad
\cos x=\frac{1-\tau^2}{1+\tau^2}.
\end{align}
For the function $f_1$, we get
\begin{align} \label{eq:f1}
f_1(\varphi)&=\frac{2A_1}{\tilde{K}}\int \frac{1}{C_1+C_2\tau^2} {\rm d}\tau\notag\\
&=\frac{2A_1}{\tilde{K}}\frac{\arctan \left(\sqrt{\frac{C_2 }{C_1}}\tau\right)}{\sqrt{C_1C_2}},
\end{align}
where we have set (cf. Eq. \eqref{eq:B1_B2})
\begin{align}
C_1&=B_1+B_2\ge0,\qquad C_2=B_1-B_2\ge0.
\end{align}
For $f_2$, we have the following result:
\begin{align} \label{eq:f2}
f_2(\varphi)&=\frac{2A_2}{\tilde{K}}\int \frac{1+\tau^2}{(C_1+C_2\tau^2)^2} {\rm d}\tau\notag\\
&=\frac{2A_2}{\tilde{K}}\biggr{[}\frac{C_1+C_2}{2 (C_1C_2)^{3/2} } \arctan\left(\sqrt{\frac{C_2}{C_1}} \tau\right)\notag\\
&-\frac{\tau (C_1-C_2)}{2 C_1 C_2 \left(C_1+C_2 \tau^2\right)}\biggr{]}.
\end{align}
Finally, the expression of $f_3$ is given by
\begin{align} \label{eq:f3}
f_3(\varphi)&=\frac{2A_3}{\tilde{K}}\int \frac{1-\tau^4}{(C_1+C_2\tau^2)^3} {\rm d}\tau\notag\\
&=\frac{2A_3}{\tilde{K}}\biggr{[}-\frac{3 \left(C_1^2-C_2^2\right) \arctan\left(\sqrt{\frac{C_2}{C_1}} \tau\right)}{8 (C_1C_2)^{5/2}}\notag\\
&+\frac{\tau C_1(3 C_1^2+5C_2^2)+(5 C_1^2+3 C_2^2)C_2\tau^3}{8 C_1^2 C_2^2 \left(C_1+C_2 \tau^2\right)^2}\biggr{]}.
\end{align} 
\begin{figure}
    \centering
    \includegraphics[scale=0.29]{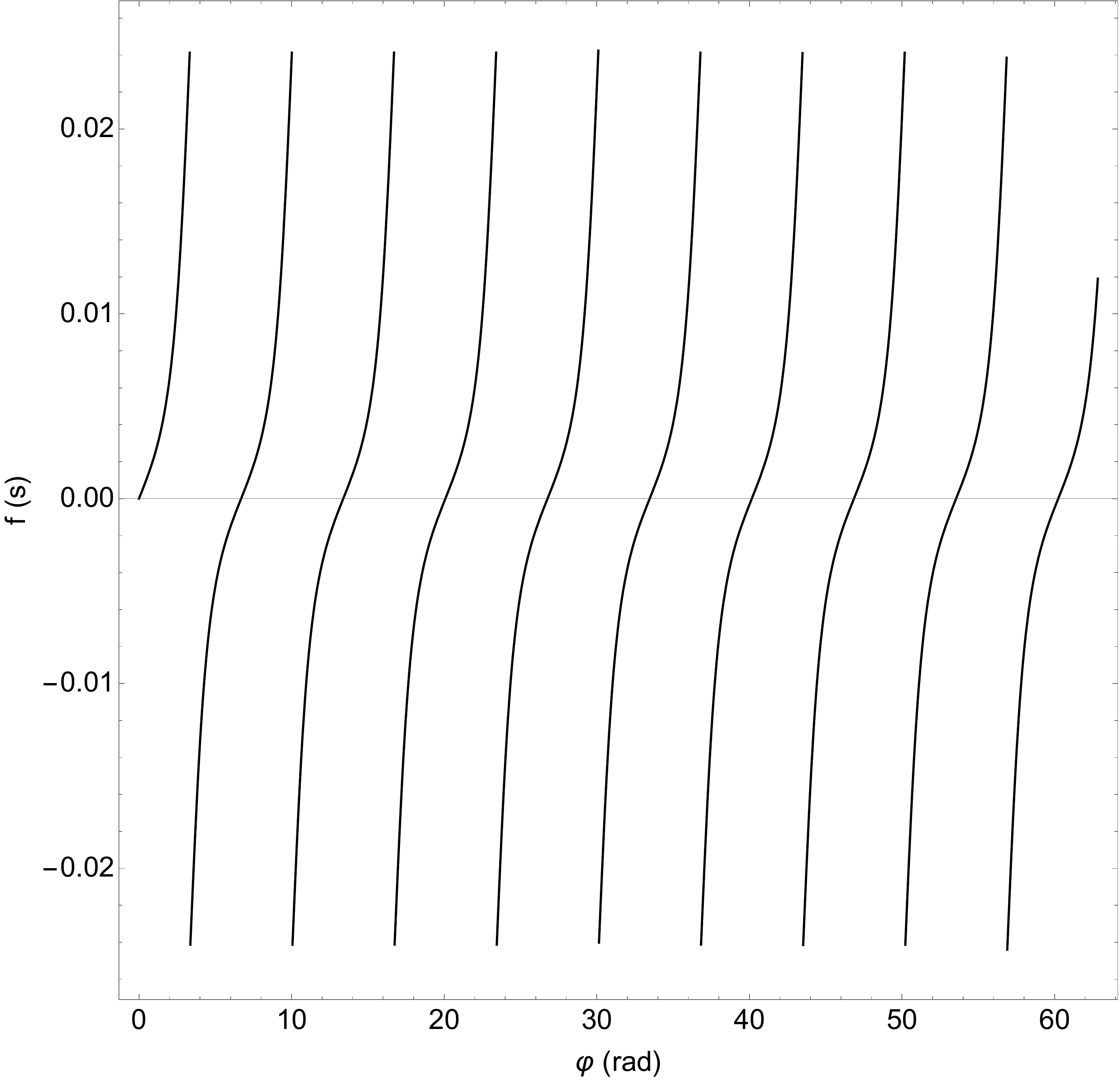}
    \caption{Function $f(\varphi)$  for $\varphi\in[0,20\pi]$, where the same numerical values as in Fig. \ref{fig:FigA1} have been used.}
    \label{fig:FigA2}
\end{figure}
The plot of the function $f(\varphi):=f_1(\varphi)+f_2(\varphi)+f_3(\varphi)$ is shown in Fig. \ref{fig:FigA2}. It is clear that $f(\varphi)$ is discontinuous, since its different periodic branches are not smoothly connected to each other; furthermore, it does not reproduce the behavior of Fig. \ref{fig:FigA1}.   
Therefore, we have to add to  it an \qm{\emph{accumulation function}}, which takes into account that time is a monotonically increasing function, while trigonometric functions and their related information are reset after each period. To this end, we define the following \emph{characteristic period}
\begin{align}
P_\varphi:=\frac{\pi}{\tilde{K}},    
\end{align}
and then the accumulation function reads as
\begin{align} \label{eq:acc_func}
F_n(\varphi):=\begin{cases}
0, & {\rm if}\ \bar{\varphi}\in[0,P_\varphi],\\
2n f(P_\varphi), & {\rm if}\ \bar{\varphi}\in[P_\varphi(2n+1),P_\varphi(2n+2)],
\end{cases}
\end{align}
where $n\in \mathbb{N}$. For a generic $\varphi$, the related value of $n$ can be calculated considering $q := [(\bar{\varphi}-P_\varphi)/P_\varphi]$, where $[\cdot]$ stands for the integer part of a number. Thus, if $q$ is an even number, then $n=(q+2)/2$; on the other hand, if $q$ is an odd number, then $n=(q+1)/2$. Therefore, we can conclude that the correct \emph{analytical} form of $t(\varphi)$ is
\begin{align} \label{eq:anal_sol}
t(\varphi)=f(\varphi)+F_n(\varphi) +{\rm O}\left(c^{-4}\right).    
\end{align}
Figure \ref{fig:FigA3} shows the agreement between the numerical solution and the analytical expression \eqref{eq:anal_sol}.
\\

\section{Discussion and conclusions} The formula for the coordinate time determined here, see Eq. \eqref{eq:anal_sol}, represents a new analytical result in the GR PN literature. The strength of our method relies upon two fundamental steps: (1) computing integrals (cf. Eqs. \eqref{eq:f1}, \eqref{eq:f2}, and \eqref{eq:f3}), which can be easily performed via a symbolic program; (2) tune the determined solution via an accumulation function (cf. Eq. \eqref{eq:acc_func}), which makes the function of point (1) continuous. The latter procedure constitutes the original aspect of the proposed analytical strategy.

At the 1PN level, this formula can be used to replace numerical schemes due to its simpler and faster implementation with no approximation costs. Importantly, our method can be generalized and extended to higher PN orders, where an analytical formula (in whatever form is presented) may improve speed and accuracy. 

We stress that the variable $\varphi$ naturally stems from the PN equations of motion for conservative binary systems. In fact, these are usually given as second-order differential equations of position with respect to the time $t$. Due to the conservation of energy and angular momentum, they can be divided into two coupled first-order ordinary differential equations of the radius $R$ and the angle $\varphi$ with respect to the time $t$. Then, it is possible to obtain an orbital equation, where we derive $R$ with respect to $\varphi$. This normally allows us to determine $R(\varphi)$, rather than $R(t)$. This constitutes a very fundamental step that then permits to find out $t(\varphi)$. We would like to underline again that the angle $\varphi$ assumes a fundamental role in dealing with the two-bodies' equations of motion both in Newtonian physics and in GR. Indeed, it replaces the time $t$ and in this way all dynamical quantities can be expressed in terms of $\varphi$, offering thus the possibility to achieve analytical results.

It is worth noting that pulsar timing recipes often make use of the inverse of Eq. \eqref{eq:anal_sol}, i.e., the function $\varphi(t)$, whose calculation still requires a numerical inversion. Here, our simplified solution can be exploited as a mean of speeding up such computations. Our finding can be generally applied for timing the orbital period of compact binary systems during the inspiral phase \cite{Postnov2014}. In addition, this formula could be extremely useful for extracting fundamental information from the \emph{holy grail binary system}, constituted by a pulsar and a black hole, as well as for providing significant tests of gravity \cite{Chattopadhyay2021}.

\acknowledgements
V.D.F. and E.B. are grateful to Gruppo Nazionale di Fisica Matematica of Istituto Nazionale di Alta Matematica for support, to Caterina Tiburzi for continuous help, and to Luigi Stella for illuminating discussions. The authors thank Alessandro Ridolfi for useful suggestions. VDF acknowledges the support of INFN {\it sez. di Napoli}, {\it iniziative specifiche} TEONGRAV. E.B. acknowledges the support of the Austrian Science Fund (FWF) grant P32086. J.A. acknowledges the support of the Stavros Niarchos Foundation (SNF) and the Hellenic Foundation for Research and Innovation (HFRI) under the 2nd Call of ``Science and Society -- Action Always strive for excellence -- ``Theodoros Papazoglou'' (Project Number: 01431).

\bibliographystyle{eplbib}
\bibliography{references}

\begin{thebibliography}{10}
\expandafter\ifx\csname url\endcsname\relax\def\url#1{\texttt{#1}}\fi

\bibitem{Maggiore:GWs_Vol1}
\Name{Maggiore M.} \Book{{Gravitational Waves. Vol. 1: Theory and Experiments}}
  Oxford Master Series in Physics (Oxford University Press) 2007.

\bibitem{Blanchet2014}
\Name{Blanchet L.} \REVIEW{Living Reviews in Relativity}{17}{2014}{2}.
\newline\url{https://doi.org/10.12942/lrr-2014-2}

\bibitem{Poisson-Will2014}
\Name{Poisson E. \and Will C.~M.} \Book{Gravity: Newtonian, Post-Newtonian,
  Relativistic} (Cambridge University Press) 2014.

\bibitem{Lorentz1937}
\Name{Lorentz H.~A. \and Droste J.} \Book{The Motion of a System of Bodies
  under the Influence of their Mutual Attraction, According to Einstein's
  Theory} (Springer Netherlands, Dordrecht) 1937 pp. 330--355.
\newline\url{https://doi.org/10.1007/978-94-015-3445-1\_11}

\bibitem{Taylor:1989sw}
\Name{Taylor J.~H. \and Weisberg J.~M.} \REVIEW{The Astrophysical
  Journal}{345}{1989}{434}.

\bibitem{Damour:1990wz}
\Name{Damour T. \and Taylor J.~H.} \REVIEW{The Astrophysical
  Journal}{366}{1991}{501}.

\bibitem{Kramer:2006nb}
\Name{Kramer M., Stairs I.~H. \etal} \REVIEW{Science}{314}{2006}{97}.

\bibitem{Kramer:2021jcw}
\Name{Kramer M., Stairs I.~H. \etal} \REVIEW{Physical Review
  X}{11}{2021}{041050}.

\bibitem{Wex:2014abc}
\Name{Wex N.} \Book{Testing {{Relativistic Gravity}} with {{Radio Pulsars}}}
  (Feb. 2014).

\bibitem{Ozel:2016oaf}
\Name{{\"O}zel F. \and Freire P.} \REVIEW{Annual Review of Astronomy and
  Astrophysics}{54}{2016}{401}.

\bibitem{Damour1985}
\Name{{Damour} T. \and {Deruelle} N.} \REVIEW{Ann. Inst. Henri Poincar{\'e}
  Phys. Th{\'e}or}{43}{1985}{107}.

\bibitem{Paper2}
\Name{Battista E. \and De~Falco V.} \REVIEW{Eur. Phys. J. C}{82}{2022}{628}.

\bibitem{Edwards:2006zg}
\Name{Edwards R.~T., Hobbs G.~B. \and Manchester R.~N.} \REVIEW{Monthly Notices
  of the Royal Astronomical Society}{372}{2006}{1549}.

\bibitem{Luo:2020ksx}
\Name{Luo J., Ransom S., Demorest P., Ray P.~S., Archibald A., Kerr M.,
  Jennings R.~J., Bachetti M., {van Haasteren} R., Champagne C.~A., Colen J.,
  Phillips C., Zimmerman J., Stovall K., Lam M.~T. \and Jenet F.~A.}
  \REVIEW{The Astrophysical Journal}{911}{2021}{45}.

\bibitem{lentali2018}
\Name{{Lentati} L., {Champion} D.~J., {Kramer} M., {Barr} E. \and {Torne} P.}
  \REVIEW{MNRAS}{473}{2018}{5026}.

\bibitem{freire2018}
\Name{{Freire} P. C.~C. \and {Ridolfi} A.} \REVIEW{MNRAS}{476}{2018}{4794}.

\bibitem{Postnov2014}
\Name{{Postnov} K.~A. \and {Yungelson} L.~R.} \REVIEW{Living Reviews in
  Relativity}{17}{2014}{3}.

\bibitem{Chattopadhyay2021}
\Name{{Chattopadhyay} D., {Stevenson} S., {Hurley} J.~R., {Bailes} M. \and
  {Broekgaarden} F.} \REVIEW{MNRAS}{504}{2021}{3682}.

\end{thebibliography}

\end{document}